**Bond-Switching Reconstruction in Bare $Ti_3C_2$ MXene**

Kui Lin*

*School of civil engineering, Sun yat-sen University*, Guangzhou, Guangdong 510275, China

*Corresponding author: Kui Lin. E-mail: linkui66@mail.sysu.edu.cn

**Abstract:** MXene properties are typically tuned through composition, surface termination, and intercalation, while the metal–carbon framework topology is considered fixed. Here, we demonstrate that this framework can reconstruct. Unconstrained relaxations of biaxially strained bare $Ti_3C_2$ yield a dynamically stable polymorph, N, featuring a 15-atom primitive cell, six distinct Ti–C bond classes, and 2.55 Å Ti–Ti pairs. The coherent transformation pathway crosses an 8.47 eV barrier per cell but first reaches another reconstructed minimum, N′, lying 1.08 eV below N, revealing a family of bond-switched networks. Phase N lies 7.58 eV per cell above the parent. Reverse barriers along the calculated pathway are 1.50 and 1.98 eV per cell for N and N′, respectively. Reconstruction quenches the parent spin polarization, yielding a nonmagnetic metal with a nearly fourfold increase in Fermi-level density of states. Bader and electron-localization analyses suggest charge redistribution from C toward Ti associated with Ti–Ti pairing. In multilayers, reconstructed layers form interlayer Ti–C bonds. A 10% biaxial tension halves the phase-energy difference, and pressure reverses phase ordering near 85 GPa. These results establish bond-switching reconstruction as a mechanically controllable structural degree of freedom in MXenes.

Different polymorphs of the same compound can exhibit distinct physical properties [1-3]. In two-dimensional (2D) materials, strain offers a means of accessing this structural diversity by shifting the relative stability of competing phases and stabilizing electronic states unavailable in the equilibrium structure [4,5]. Studies of polymorphic transformations in 2D transition-metal dichalcogenides have explored pathways involving collective lattice distortions and atomic-plane gliding [6,7]. Reconstructive transformations, by contrast, require extensive bond breaking and reformation and remain less widely explored. Such transformations extend beyond distortions of an existing lattice. They alter the connectivity of the bonding network itself, potentially creating metastable structures with electronic and magnetic properties qualitatively different from those of the parent phase.

MXenes, a family of 2D transition-metal carbides and nitrides derived from MAX-phase precursors [8,9], provide a compelling platform for investigating this possibility. They combine metallic conduction with tunable surface chemistry and diverse electronic states [10], while several bare, termination-free members are predicted to exhibit magnetism [11,12]. To date, property engineering in MXenes has relied predominantly on composition, surface termination, defects, intercalation, and stacking. Although these approaches modify local bonding and electronic structure, the connectivity of the underlying metal–carbon framework is generally treated as a fixed structural constraint rather than an independent design variable. Whether this framework can reconstruct into a distinct, metastable bonding network at fixed stoichiometry has, to our knowledge, remained unresolved. Addressing this question requires establishing not only the existence of alternative structures but also their dynamical stability, transformation pathways, and electronic consequences.

Here, we identify a reconstructive transformation in bare $Ti_3C_2$ through unconstrained structural relaxations under biaxial strain. The resulting polymorph, N, combines strongly differentiated Ti–C coordination with short Ti–Ti pairs, an in-plane area 50% larger than that of the parent, and a layer thickness 35% smaller. The coherent transformation pathway first reaches a second reconstructed minimum, N′, lying below N in energy, rather than reaching N directly. Both structures are dynamically stable and separated from the parent by reverse barriers along this pathway. Reconstruction thus produces a family of metastable, bond-switched networks rather than an isolated alternative structure.

Reconstruction quenches the parent spin polarization and enhances the Fermi-level density of states, yielding a nonmagnetic metal with local Ti–Ti pairing. In multilayers, the reconstructed framework enables interlayer Ti–C bonding and makes phase competition sensitive to tension and pressure. These findings establish bond connectivity as a tunable structural degree of freedom in MXenes and identify framework reconstruction as a route to mechanically control electronic and magnetic states at fixed composition.

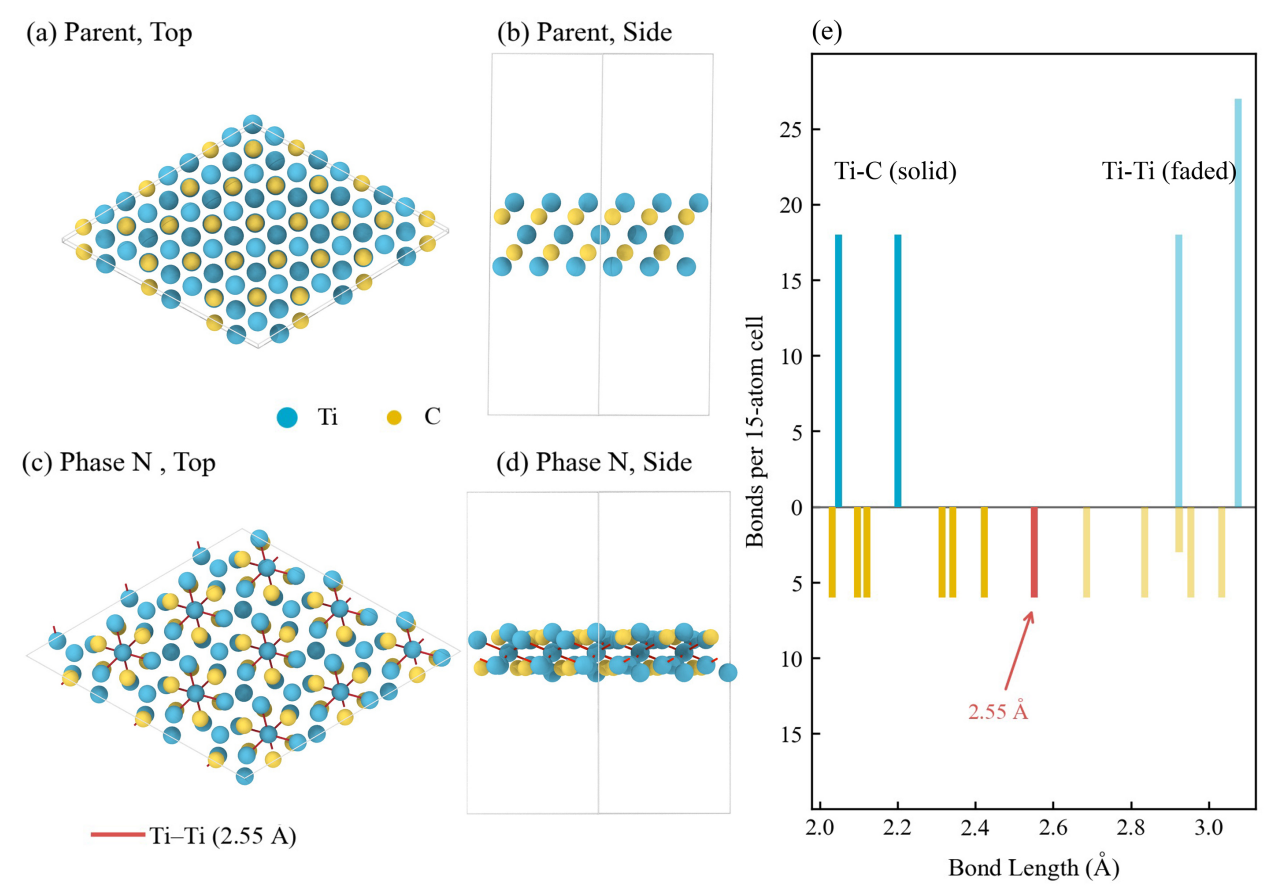

**Fig. 1. Structures and characteristic interatomic distances of the parent and phase-N $Ti_3C_2$ monolayers.** (a), (b) Top and side views of the parent structure in the 15-atom reference supercell. (c), (d) Corresponding views of phase N in its primitive 15-atom cell. Ti is light blue, C dark gray; short Ti–Ti pairs (2.55 Å) are highlighted in red. Cell boundaries are indicated by dashed lines. (e) Ti–C (opaque) and Ti–Ti (faded) distance distributions of the two phases, plotted upward for the parent and downward for phase N.

**Discovery and structure.** Starting from parent monolayers initialized at biaxial strains up to 34%, we relaxed the cell and atomic positions without imposing symmetry. Initial strains of 33% and 34% drove the structures out of the parent basin, yielding a fully relaxed, commensurate 15-atom phase, denoted N (Fig. 1). The parent has a five-atom primitive cell with $P\bar{3}m1$ symmetry, Ti–C–Ti–C–Ti stacking, two Ti–C bond classes at 2.05 and 2.20 Å, and a nearest Ti–Ti distance of 2.92 Å [Figs. 1(a) and 1(b)]. Phase N derives from a $\sqrt{3} \times \sqrt{3}\, R30°$ supercell of the parent, folding the parent $K$ point onto Γ [Figs. 1(c) and 1(d)]. Its relaxed symmetry is $P\bar{1}$(No. 2), although all atoms lie within 0.01 Å of positions compatible with $P\bar{3}$ symmetry (No. 147).

For equivalent 15-atom cells, reconstruction increases the in-plane area by 50%, from 24.57 to 36.88 Å², while reducing the layer thickness by 35%, from 4.65 to 3.02 Å. These changes nearly compensate, leaving a geometric slab-volume change of only −2.5%. The local bonding, by contrast, is extensively reorganized [Fig. 1(e)]. The two parent Ti–C bond classes split into six, spanning 2.03–2.42 Å. Selected Ti–Ti distances contract from 2.92 to 2.55 Å, forming short pairs, while the remaining separations extend up to 3.03 Å. The Ti sites develop distinct local environments: undercoordinated sites with three short Ti–C bonds, sites forming short Ti–Ti pairs, weakly bonded sixfold-coordinated sites, and intermediate environments. This reorganization of bond connectivity identifies the transformation as reconstructive rather than displacive. Phase N is dynamically stable. Its phonon dispersion, computed by the finite-displacement method [13], shows no physically significant imaginary branch [Fig. 2(a)].

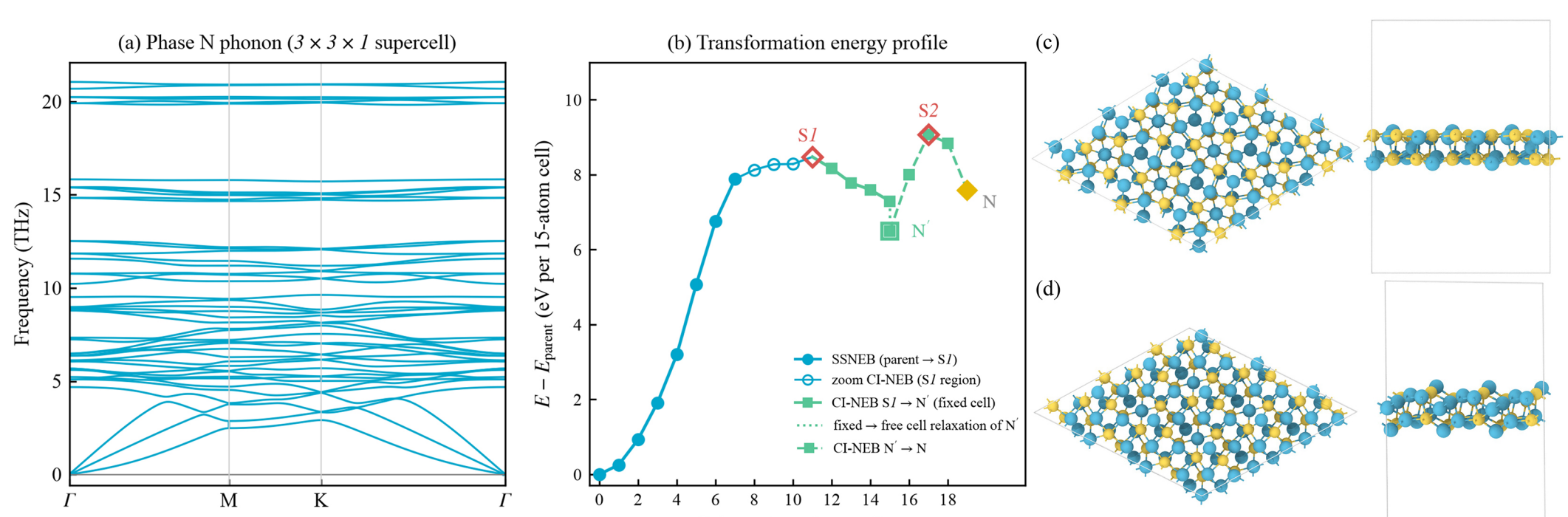

**Fig. 2. Dynamical stability and resolved coherent transformation landscape.** (a) Phonon dispersion of phase N calculated using a $3 \times 3 \times 1$ supercell along $\Gamma{-}M{-}K{-}\Gamma$.Their coordinates are given in the Supplemental Material (SM). (b) Resolved prescribed-cell transformation profile connecting the parent, primary path maximum

$S_1$, reconstructed minimum N′, secondary path maximum $S_2$, and phase N. Non-spin-polarized PBE-D3 energies are reported per 15-atom cell relative to the relaxed nonmagnetic parent. Open circles denote the converged, locally refined climbing-image NEB (CI-NEB) calculation around $S_1$. The fixed-cell endpoint on the N′ branch lies 0.79 eV above fully relaxed N′ and relaxes to it upon release of the cell constraints; the dotted segment indicates this relaxation [SM, Sec. IV.A and Table S6]. The dashed N′→N segment denotes a CI-NEB band with three intermediate images. Connecting lines are guides to the eye and do not imply additional computed configurations. (c), (d) Top and side views of the $3 \times 3 \times 1$ supercells used for the phonon calculations of N and N′, respectively.

**Transformation landscape.** We mapped the coherent transformation using a prescribed-cell NEB calculation with eight intermediate images [14-16]. Lattice vectors were linearly interpolated between the endpoints and fixed at each image, while atomic positions were optimized under NEB forces; unlike generalized solid-state NEB, the cell and atomic coordinates were not jointly optimized [15]. The primary path maximum, $S_1$, lies 8.47 eV per 15-atom cell above the nonmagnetic parent. Referencing the same non-spin-polarized maximum to the spin-polarized parent ground state gives an energy difference of 8.73 eV. The large energy cost accompanies extensive bond switching along the prescribed coherent route.

Beyond $S_1$, the refined band terminates at a fixed-cell structure that relaxes to a second reconstructed minimum, N′, upon release of the cell constraints, lowering the energy by 0.79 eV [Fig. 2(b); SM, Sec. IV.A and Table S6]. Fully relaxed N′ lies 1.08 eV below N and retains its characteristic short Ti–Ti pairs, with separations of 2.52–2.59 Å and differentiated Ti–C coordination (Fig. S2). Relative to N, its sheared $P\bar{1}$cell has a 9% smaller in-plane area and a thicker layer of 4.16 Å. Phonon calculations also support its dynamical stability (Fig. S3). A separate CI-NEB band connects N′ to N through the secondary path maximum $S_2$, 9.07 eV above the parent. The two minima identify a family of metastable $Ti_3C_2$ networks assembled from similar local bonding motifs.

Energy differences from the fully relaxed minima are 1.98 eV from N′ to $S_1$, 1.50 eV from N to $S_2$, and 2.57 eV from N′ to $S_2$. The N–$S_1$ energy difference is not an escape barrier, because the resolved route from N first crosses $S_2$. These values describe an assembled prescribed-cell profile, including relaxation of the N′ endpoint, rather than a variable-cell minimum-energy path; they neither establish lifetimes nor exclude lower-barrier routes involving defects, edges, or interphase boundaries.

For the isolated monolayer, N and N′ lie 7.58 and 6.50 eV per cell above the nonmagnetic parent, respectively. Across the tested computational settings, single-point calculations at the existing pathway geometries give N–parent energy differences of 6.51–7.92 eV and primary path-maximum energies of 7.94–8.91 eV (Tables S2–S4). These tests support the qualitative energetic picture but do not constitute independently optimized pathways.

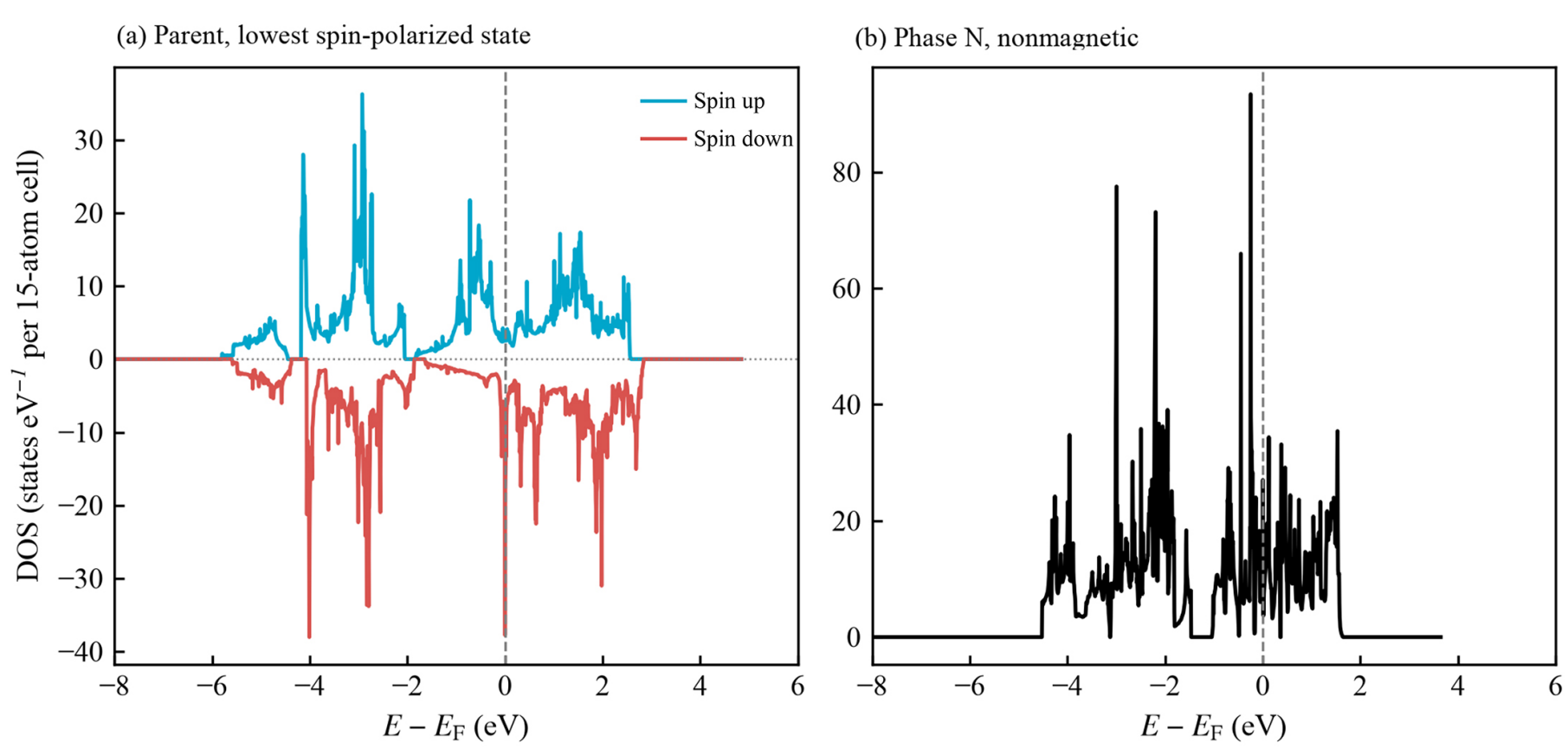

**Fig. 3. Electronic densities of states of the parent and reconstructed monolayers.** DOS values are reported in states $eV^{-1}$ per 15-atom cell, with the Fermi level set to zero. (a) Spin-resolved DOS of the lowest-energy parent state found among the tested magnetic configurations; the spin-down DOS is plotted with a negative sign. (b) Total DOS of phase N.

**Magnetic quenching and enhanced metallicity.** The parent monolayer is a spin-polarized metal [Fig. 3(a) and Fig. S4]. Among the tested magnetic configurations, the lowest-energy parent state has a total moment of 5.37 μB per 15-atom cell, dominated by parallel moments on the outer Ti atoms, and lies 0.26 eV below the nonmagnetic solution (Table S1). Both spin channels cross the Fermi level, consistent with previous predictions for bare $Ti_3C_2$ [11,12].

Reconstruction strongly suppresses the spin polarization. All tested magnetic initializations for N converge to a low-moment solution, with a total moment of approximately 0.2 μB per cell and an energy spread below 0.02 meV. Fixed-spin-moment calculations show increasing energy penalties over the tested range of 1–8 μB (SM). Nevertheless, N remains metallic, with several bands crossing $E_{\mathrm{F}}$(Fig. S5) and $N(E_{\mathrm{F}}) \approx 11.3$states eV⁻¹ per cell [Fig. 3(b)]. A spin-polarized calculation at the geometry of the primary path maximum $S_1$yields a reduced moment of 0.85 μB, linking the magnetic suppression to the structural reorganization along the prescribed route.

**Local rebonding.** The commensurate superstructure, bond disproportionation, and short Ti–Ti pairs invite comparison with charge-density-wave (CDW) distortions [17], including those predicted in nitride and chalcogenide MXene-type compounds [18-20]. Here, however, reconstruction increases rather than depletes the DOS at the Fermi level relative to the chosen structural reference: an ideal parent constrained to the phase-N lattice has $N(E_{\mathrm{F}}) \approx 3.0$states eV⁻¹ per 15-atom cell, compared with 11.3 states eV⁻¹ for N, a 3.8-fold increase (Fig. S6). This comparison does not support a simple DOS-depletion picture, although it does not by itself exclude a CDW-related mechanism.

Bader analysis [21] indicates a net redistribution of approximately 1.6 electrons per cell from C to Ti upon reconstruction (Table S7). Ti electron populations vary with coordination, with the largest values at the short-paired sites. The electron localization function [8] [22] also changes at the short Ti–Ti contacts (Fig. S7). At the contact midpoint, it increases from 0.11 in the parent at 2.92 Å to 0.29 in N at 2.55 Å and 0.32 in N′ at 2.52 Å. Together, the bond contraction, charge redistribution, and increased ELF support local rebonding at the short Ti–Ti contacts, but do not establish preferential occupation of Ti–Ti bonding states. Preferential occupation of Ti–Ti bonding states is a possible microscopic explanation but is not established by the present charge and ELF analyses alone.

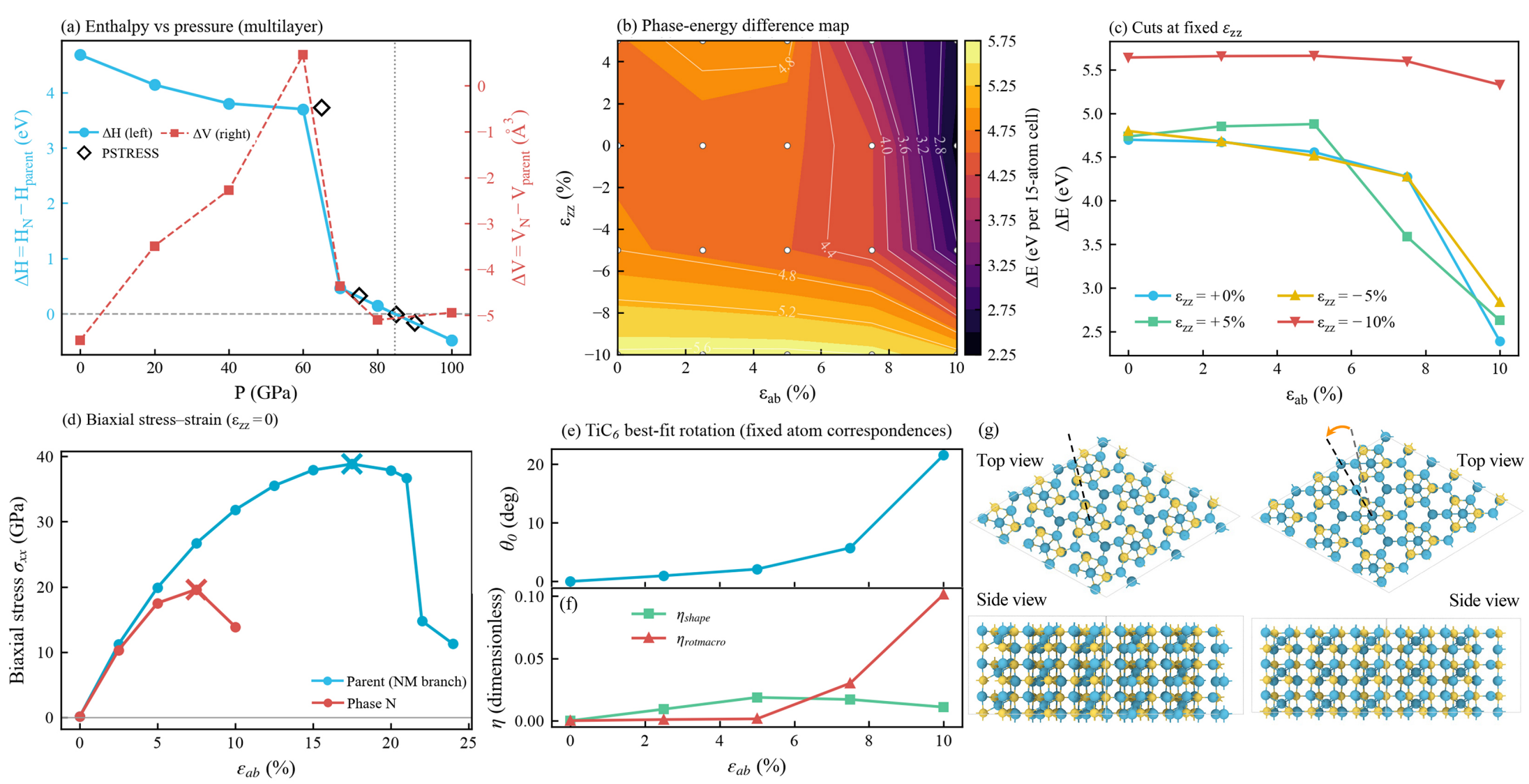

**Fig. 4. Strain- and pressure-dependent energetics and local $TiC_6$-unit reorientation in periodic multilayer cells.** (a) Enthalpy difference $\Delta H(P) = H_{\mathrm{N}} - H_{\mathrm{parent}}$ (left axis) and volume difference $\Delta V(P) = V_{\mathrm{N}} - V_{\mathrm{parent}}$(right axis). Symbols denote calculated points. $P^* \approx 85$ GPa marks the estimated enthalpy crossover within the two-structure, fixed-stoichiometry comparison. Black open diamonds are an independent constant-pressure (PSTRESS) relaxation series confirming the crossover. (b) Energy difference $\Delta E = E_{\mathrm{N}} - E_{\mathrm{parent}}$ under combined in-plane biaxial strain $\varepsilon_{ab}$and out-of-plane strain $\varepsilon_{zz}$. Symbols mark calculated points, and contours are interpolated between them. (c) Cuts through (b) at fixed $\varepsilon_{zz}$. (d) Biaxial stress–strain responses at $\varepsilon_{zz} = 0$, plotted as $\sigma_{xx}$. Stresses are the three-dimensional stresses of the periodic multilayer cells, with tension taken as positive. In (b)–(d), strains are referenced to each phase's own equilibrium cell. (e) Best-fit rotation angle $\theta_0$ of the analyzed $TiC_6$ unit relative to its zero-strain configuration. (f) Normalized fitting residuals after rotation plus uniform scaling ($\eta_{\mathrm{shape}}$) and after macroscopic affine deformation followed by an additional optimal rotation ($\eta_{\mathrm{rotmacro}}$). Fits retain fixed atom correspondences, and residuals are normalized by the reference root-mean-square Ti–C bond length. (g) Top and side views of the N-derived multilayer at 7.5% and 10% biaxial strain, illustrating the reorientation of the analyzed $TiC_6$ unit.

**Interlayer bonding in multilayers.** To examine the effect of stacking, we next considered vacuum-free periodic multilayer cells. The two phases develop markedly different interlayer geometries (Fig. S9). The parent retains a 2.47 Å interlayer gap, with nearest interlayer Ti–Ti contacts of 3.03 Å. The N-derived multilayer instead forms six interlayer Ti–C bonds of 2.14 Å per cell, one involving each surface Ti atom, connecting the layers into a network that extends along the stacking direction. The per-layer period decreases from 7.15 to 4.43 Å, while the in-plane area increases by 56%, giving a net volume reduction of approximately 3.2%. The short Ti–Ti pairs survive, and each surface Ti atom retains three C neighbors, one of which now belongs to the adjacent layer (Table S11).

Assembly into the multilayer provides 2.88 eV per cell more stabilization for N than for the parent, reducing the energy difference from 7.58 eV in the monolayer to 4.70 eV in the multilayer. The pathway barriers discussed above refer to the prescribed monolayer cell; the loading responses below instead characterize the periodic multilayer structures, including the interlayer-connected N-derived network.

**Mechanical response and local reorientation.** We examined in-plane biaxial strain, out-of-plane strain, shear, and hydrostatic pressure (Fig. 4). Among the sampled strain modes, in-plane tension produces the largest reduction in $\Delta E$: at 10% biaxial strain and $\varepsilon_{zz} = 0$, the difference decreases from 4.70 to 2.39 eV per cell [Figs. 4(b) and 4(c)]. Out-of-plane compression has the opposite effect. At $\varepsilon_{zz} = -10\%$ and zero in-plane strain, $\Delta E$ rises to 5.64 eV, consistent with the parent accommodating compression through its interlayer gap. Pure shear produces only a modest reduction in $\Delta E$ (Fig. S8). Because strains are referenced to each phase's own equilibrium cell, these energy differences compare responses to equal nominal strains rather than structures constrained to a common lattice metric.

At $\varepsilon_{zz} = 0$, the parent has an initial biaxial modulus of approximately 450 GPa and reaches a stress maximum of about 39 GPa near 17.5–20% strain [Fig. 4(d)]. The N-derived structure has an initial modulus of approximately 410 GPa and reaches about 20 GPa near 7.5% strain, followed by a stress drop. Between 7.5% and 10% strain, the relaxed N-derived energy decreases by 0.14 eV. The accelerated decrease in $\Delta E$ over this interval therefore accompanies a substantial structural rearrangement, while the short Ti–Ti pairs remain intact.

To characterize this rearrangement, we tracked one $TiC_6$ unit using fixed atom correspondences [Figs. 4(e)–4(g)]. Its best-fit rotation increases from 0.97° at 2.5% strain to 21.55° at 10%. At 10% strain, a rotation-plus-uniform-scaling fit gives a normalized residual of 0.011 and a scale factor of 0.992, whereas a fit based on macroscopic affine deformation followed by optimal rotation leaves a residual of 0.101. The analyzed unit thus undergoes substantial nonaffine reorientation with little change in size or shape, consistent with a local nearly rigid-unit response that may contribute to stress release. Establishing a collective mechanism requires analysis of the other coordination units and their connectivity.

**Pressure-induced collapse and enthalpy crossover.** Hydrostatic pressure provides a complementary comparison at a common imposed pressure [Fig. 4(a)]. The enthalpy penalty decreases from 4.68 eV per cell at 0 GPa to 3.70 eV at 60 GPa. Between 60 and 70 GPa, the N-derived structure undergoes a collapse. At 65 GPa, compression and decompression scans yield two distinct solutions whose enthalpies differ by 3.12 eV. This history dependence is consistent with a first-order-like collapse and metastability, although it does not locate the equilibrium transition pressure between the uncollapsed and collapsed states.

The collapsed N-derived branch is approximately 3–4% denser than the parent and retains short Ti–Ti pairs, with a separation of 2.25 Å at 85 GPa. Its enthalpy crosses that of the parent at an estimated $P^* \approx 85$GPa (Table S9). This crossover applies only to the two compared structural families at fixed stoichiometry. It does not establish global thermodynamic stability because alternative structures and decomposition products were not considered.

Whether the reconstructed network can be generated along an accessible pathway and retained after unloading remains unresolved.

**Role of surface terminations.** To assess the sensitivity to surface chemistry, we compared fully O-terminated $Ti_3C_2O_2$ structures derived from both phases (SM, Sec. VII). In the tested N-derived configuration, the short Ti–Ti network survives relaxation, with the shortest contact of 2.52 Å. However, its energy lies 5.86 eV per formula unit above the calculated O-terminated parent, compared with a bare-surface penalty of 2.53 eV per formula unit. Thus, full oxygen termination preserves the short-pair motif but makes reconstruction energetically less favorable in the configurations examined.

**Summary.** The Ti–C framework of bare $Ti_3C_2$ is not topologically unique. Strain-assisted unconstrained relaxation reveals dynamically stable, bond-switched polymorphs characterized by short Ti–Ti pairs and differentiated Ti–C coordination. The resolved coherent pathway identifies barriers separating N and N′, supporting their metastability along the examined monolayer transformation route. Reconstruction converts a spin-polarized metal into a nonmagnetic metal with an enhanced Fermi-level density of states, with these electronic changes linked to local Ti–Ti pairing. In multilayers, interlayer Ti–C bonds connect the reconstructed layers into a network, and the analyzed $TiC_6$ unit undergoes nearly rigid, nonaffine reorientation under tension. A pressure-induced collapse brings the N-derived branch below the parent in enthalpy near 85 GPa within the two-structure, fixed-stoichiometry comparison, without establishing global thermodynamic stability. Full oxygen termination increases the reconstruction penalty in the tested configuration, underscoring the importance of surface chemistry. These results identify framework topology as an additional design variable alongside composition, termination, and stacking in MXenes. More broadly, unconstrained relaxation of strongly deformed lattices may provide a computational strategy for discovering hidden bonding topologies in other strongly bonded two-dimensional materials.

**Computational details.** DFT calculations used VASP [23] with PAW potentials [24,25], PBE [26], and D3 dispersion corrections with Becke–Johnson damping [27,28]. Reconstructed structures were identified by symmetry-unconstrained relaxation of strained parent cells, releasing both cell and ionic degrees of freedom. Coherent pathways were evaluated using prescribed-cell NEB: each image's cell was fixed by interpolation between the endpoints, while atomic positions were optimized. The reported barriers thus refer to this constrained route. Phonons were calculated using PHONOPY [13]. Computational parameters, convergence tests, electronic and bonding analyses, and additional functional and PAW-potential checks are detailed in the Supplemental Material.

**Acknowledgments**

This work was supported by the Science and Technology Program of Guangdong Province (Grant No. 2024QN11G117) and the start-up funding from Sun Yat-Sen University to Kui Lin.

**Data availability.** The data supporting the findings of this study are provided in the Supplemental Material. Additional raw calculation outputs and analysis scripts are available from the corresponding author upon reasonable request.

**References**

[1] J. Lim, J.-I. Lee, Y. Wang, N. Gauriot, E. Sebastian, M. Chhowalla, C. Schnedermann, and A. Rao, Nature **633**, 83 (2024).

[2] L. Mei *et al.*, Nature Synthesis **4**, 303 (2025).

[3] D. W. Davies *et al.*, Nature Communications **14**, 1304 (2023).

[4] K.-A. N. Duerloo, Y. Li, and E. J. Reed, Nature communications **5**, 4214 (2014).

[5] T. Li, S. Deng, H. Liu, and J. Chen, Chem. Rev. **124**, 7045 (2024).

[6] M.-X. Guan, X.-B. Liu, D.-Q. Chen, X.-Y. Li, Y.-P. Qi, Q. Yang, P.-W. You, and S. Meng, Phys. Rev. Lett. **128**, 015702 (2022).

[7] M. Chaudhary *et al.*, Mater. Today **84**, 28 (2025).

[8] M. Naguib, M. Kurtoglu, V. Presser, J. Lu, J. Niu, M. Heon, L. Hultman, Y. Gogotsi, and M. W. Barsoum, Adv. Mater. **23**, 4248 (2011).

[9] M. Naguib, O. Mashtalir, J. Carle, V. Presser, J. Lu, L. Hultman, Y. Gogotsi, and M. W. Barsoum, ACS nano **6**, 1322 (2012).

[10] B. Anasori, M. R. Lukatskaya, and Y. Gogotsi, Nature Reviews Materials **2**, 16098 (2017).

[11] M. Khazaei, M. Arai, T. Sasaki, C. Y. Chung, N. S. Venkataramanan, M. Estili, Y. Sakka, and Y. Kawazoe, Adv. Funct. Mater. **23**, 2185 (2013).

[12] Y. Xie and P. R. Kent, Physical Review B—Condensed Matter and Materials Physics **87**, 235441 (2013).

[13] A. Togo and I. Tanaka, Scripta Mater. **108**, 1 (2015).

[14] D. Sheppard, R. Terrell, and G. Henkelman, The Journal of chemical physics **128** (2008).

[15] D. Sheppard, P. Xiao, W. Chemelewski, D. D. Johnson, and G. Henkelman, The Journal of chemical physics **136** (2012).

[16] G. Henkelman, B. P. Uberuaga, and H. Jónsson, The Journal of chemical physics **113**, 9901 (2000).

[17] G. Grüner, Rev. Mod. Phys. **60**, 1129 (1988).

[18] J. Seeyangnok, U. Pinsook, and G. J. Ackland, arXiv preprint arXiv:2603.06284 (2026).

[19] S.-X. Qiao, K.-Y. Jiang, Y.-L. Han, N. Jiao, Y.-J. Chen, H.-Y. Lu, and P. Zhang, Physical Review B **112**, 205408 (2025).

[20] S. X. Qiao, Y. L. Han, K. Y. Jiang, N. Jiao, C. Si, H. Y. Lu, and P. Zhang, Adv. Funct. Mater. **35**, 2502257 (2025).

[21] G. Henkelman, A. Arnaldsson, and H. Jónsson, Comp. Mater. Sci. **36**, 354 (2006).

[22] A. D. Becke and K. E. Edgecombe, The Journal of chemical physics **92**, 5397 (1990).

[23] G. Kresse and J. Furthmüller, Physical review B **54**, 11169 (1996).

[24] P. E. Blöchl, Physical review B **50**, 17953 (1994).

[25] G. Kresse and D. Joubert, Physical review b **59**, 1758 (1999).

[26] J. P. Perdew, K. Burke, and M. Ernzerhof, Phys. Rev. Lett. **77**, 3865 (1996).

[27] S. Grimme, J. Antony, S. Ehrlich, and H. Krieg, The Journal of chemical physics **132** (2010).

[28] S. Grimme, S. Ehrlich, and L. Goerigk, J. Comput. Chem. **32**, 1456 (2011).